# V-BAND OPTOELECTRONIC OSCILLATOR FOR EARTH OBSERVATION APPLICATIONS

Dimitrios Kastritsis*[a], Enrico Lia[b], Iain Mckenzie[b], and Stavros Iezekiel[a]
[a]Emphasis Research Centre, University of Cyprus, Nicosia, Cyprus; [b]European Space Research and Technology Centre, European Space Agency, Noordwijk, Netherlands

## ABSTRACT

An optoelectronic oscillator (OEO) producing a signal at 45.86 GHz is demonstrated that may potentially be utilized in the local oscillator (LO) generation of Earth observation applications such as the microwave sounding (MWS), microwave imaging (MWI) and ice cloud imaging (ICI) missions of METOP 2 of ESA. Preliminary results show that the performance of this lower SWaP OEO system is comparable or in some respects better than the electrical system already used in the MWS, MWS and MWI missions of ESA. Specifically, a sidemode suppression of about 46 dB, a phase noise of -102 dBc/Hz at 100 kHz offset, and frequency stability of 30 kHz in a ten-minute interval is achieved for the 45.86 GHz signal. The OEO is a promising candidate to replace or supplement the electronic systems of phase locked dielectric resonator oscillator and frequency multipliers used currently in Earth observation.

*Email: kastritsis.dimitrios@ucy.ac.cy

## 1. INTRODUCTION

Optoelectronic oscillators (OEO) are capable of producing low-phase-noise microwave signals with or without the need for a microwave reference signal [1] [2]. The main advantage of an OEO is that the phase noise is independent of the output frequency meaning that it is possible in principle to generate up to sub-THz signals with ultra-low phase noise [3].

OEOs can be adapted to minimize the size, weight, and power (SWaP) requirements using a semiconductor optical amplifier (SOA) instead of a power-hungry erbium, doped fiber amplifier (EDFA) and a compact fiber coil of several kilometers [4] instead of bulkier fiber spools. This implementation is suitable potentially for space applications where SWaP considerations are of utmost importance. More specifically, in the domain of Earth observation, an intermediate signal for LO generation implemented with an OEO can in principle replace or supplement the existing LO chain based on frequency multiplication and synthesis in MWS, MWI and ICI missions of METOP2 of ESA used for weather and climate monitoring [5-7]. For example, an OEO producing a signal at 45.8 GHz combined with a frequency doubler can be used as an intermediate signal for LO generation at 183 GHz, substituting a phase locked dielectric resonator oscillator and a Schottky tripler in the MWS front end receiver [7], if it is more advantageous in terms of size, power consumption and electrical performance.

In this paper, the preliminary results on the electrical performance of this lower SWaP OEO system are demonstrated with a comparison of performance metrics to the electronic system that is currently planned (phase locked dielectric resonator oscillator and frequency multipliers) to be used in the MWI and MWS radiometers. The outline of this paper is as follows: In Section 2, the OEO experimental setup and operating parameters are provided. In Section 3, the results of the OEO are demonstrated with a comparison of the electrical performance between the OEO produced signal and the intermediate electrical signal used for LO generation in Earth observation. The paper is concluded in Section 4.

## 2. EXPERIMENTAL SETUP AND OPERATING PARAMETERS

The proposed dual loop OEO is shown in Fig. 1. A tunable laser source (LS) produces an optical carrier signal at 1550.76 nm with power of +10 dBm. A variable optical attenuator (VOA) is used to adjust the power of the optical signal, which is then modulated using a quadrature-biased 40 GHz Mach-Zehnder Modulator (MZM) with $V_{\pi, DC}$ = 3.62 V. The modulated signal is subsequently amplified by a semiconductor optical amplifier (SOA). The bias current of the SOA is

570 mA and the power at its input is adjusted by the VOA so that the SOA is in saturation which facilitates OEO stable operation and has been shown to improve the phase noise performance of microwave photonic links [8]. The 10% output of a 90-10 splitter is used to monitor the optical signal at the SOA output using an optical spectrum analyzer (OSA). The optical signal is then divided into two parts by a 50-50 optical splitter, into two spools of single mode fiber (SMF), SMF1 and SMF2 with lengths of 100 m and 1 km, respectively. The optical signals at the output of SMF1 and SMF2 are converted to the electrical domain using a 43 GHz balanced photodiode (BPD). The sum of input optical power at both photodiodes of the BPD for stable operation is +4.2 dBm. The current of the photodiodes of BPD for stable operation is 0.57 mA and 0.49 mA. The electrical signal is then amplified by a 30 dB Q-band low noise amplifier (LNA) with a noise figure of 4 dB. The amplified signal is then filtered by a narrowband electrical bandpass filter (EBPF) centered at 45.8 GHz with a bandwidth of 480 MHz. Finally, one of the output ports of the 50-50 electrical splitter is used to observe the signal with the electrical spectrum analyzer (ESA), while the signal of the other output port is amplified by a 30 dB power amplifier (PA) with an output saturation power of +27 dBm before closing the loop at the MZM RF input port.

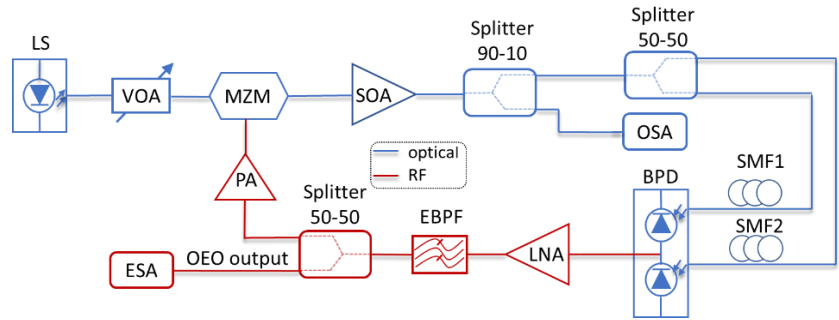

Fig. 1. Schematic diagram of the experimental OEO. LS: Laser Source. VOA: Variable Optical Attenuator. MZM: Mach-Zehnder Modulator. SOA: Semiconductor Optical Amplifier. OSA: Optical Spectrum Analyzer. SMF: Single Mode Fiber. BPD: Balanced Photodiode. LNA: Low Noise Amplifier. EBPF: Electrical Bandpass Filter. PA: Power Amplifier. ESA: Electrical Spectrum Analyzer.

For the ESA, the Keysight PNA 5227B with option S930905B Spectrum Analysis up to 50 GHz is used. Compact fiber coils were not available in our laboratory, so spools of regular size were used. The components used for the OEO implementation are shown in Table 1. A photo of the experimental bench with the OEO implementation, identifying every component, is shown in Fig. 2.

Table 1. Components used for the 45.86 GHz OEO implementation.

| Component | Abbreviation | Product |
|---|---|---|
| Laser Source | LS | Chilas CT3 C-band |
| Variable Optical Attenuator | VOA | Thorlabs VOA 50 |
| Mach-Zehnder Modulator | MZM | iXblue MXAN-LN-40-PD-P-P-FA-FA |
| Semiconductor Optical Amplifier | SOA | COVEGA BOA 1004 |
| Optical Splitter 90-10 | - | Regular 90-10 optical splitter |
| Optical Splitter 50-50 | - | Regular 50-50 optical splitter |
| Single Mode Fiber | SMF | Regular size fiber of 100 m and 1 km |
| Balanced Photodiode | BPD | Finisar BPDV2120R |
| Low Noise Amplifier | LNA | Eravant SBL-3335033040-2F-2F |
| Electrical Bandpass Filter | EBPF | IMST custom waveguide filter |
| Electrical splitter 50 -50 | - | Marki microwave PD-0465 |
| Power Amplifier | PA | Eravant SBP-4035533026-VFVF-S1 |

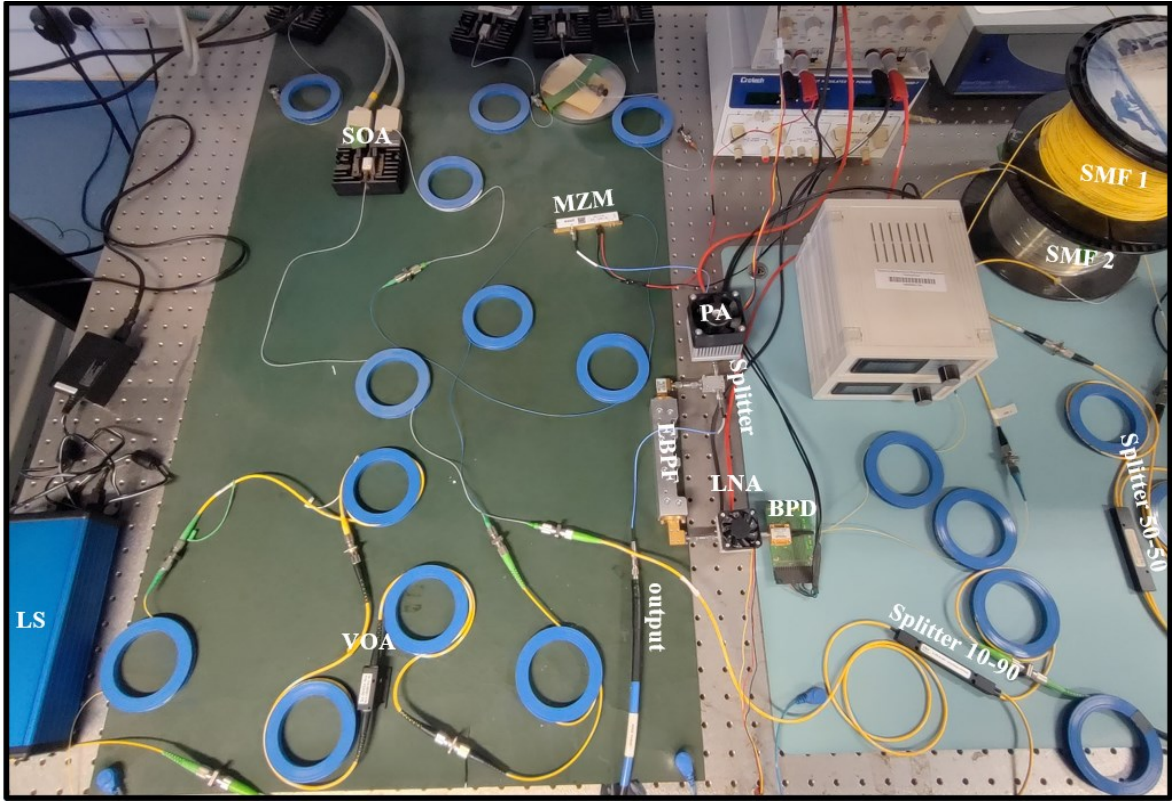

Fig. 2. Photo of the experimental bench with the OEO implementation.

## 3. RESULTS AND DISCUSSION

The oscillation frequency of the OEO is determined by a combination of the amplitude and phase response of the optical and electrical components of the loop. The EBPF plays a key role as it filters out most of the frequency components and passes a 550 MHz band around 45.8 GHz. For the combination of commercially available components and parameters that we used, the oscillation frequency of the OEO is at 45.86 GHz. The FSR of the optoelectronic oscillator is determined by the shorter loop [9] which in our case is approximately 115 m when we consider the shorter fiber spool of 100 m while adding approximately 15 m of fiber patch cords used for connecting the optical devices of Fig.1. In the case of a loop with length 115 m the FSR is 1.8 MHz. Fig. 3 shows the electrical spectrum of the output signal captured by the ESA. In Fig. 3, the frequency differences of the OEO output at 45.859 GHz and the local peaks are equal to multiples of the FSR which is approximately equal to 1.8 MHz, while the sideband suppression ratio is approximately 46 dB.

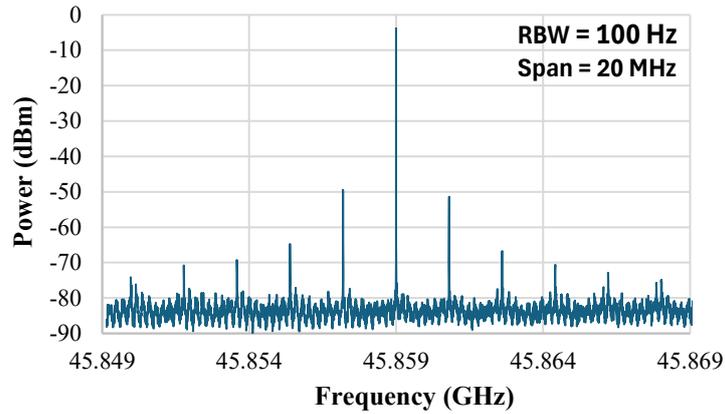

Fig. 3. Electrical spectrum captured at the OSA. The span is 20 MHz and the resolution bandwidth is 100 Hz.

Fig. 4 shows the optical spectrum captured by the OSA for a stable signal at the output. The optical carrier is double sideband modulated by a 45.86 GHz electrical signal. Due to the high RF power (~ 23 dBm) at the MZM RF input, the MZM operates in nonlinear mode, and the second harmonic at 91.72 GHz from the optical carrier is also observed at the optical spectrum.

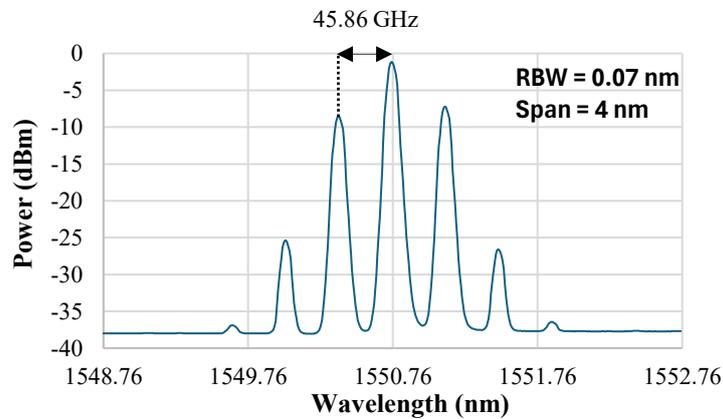

Fig. 4. Optical spectrum at OSA.

Fig. 5 shows the single sideband phase noise of the produced signal. The phase noise at offset frequencies between 1 kHz and 4 MHz was calculated from the electrical spectrum captured by the ESA. A phase noise of -90 dBc/Hz at 10 kHz and -102 dBc/Hz at 100 kHz is achieved. In the phase noise curve, the sidemodes at 1.8 MHz and at 3.6 MHz are observed due to the FSR of the OEO.

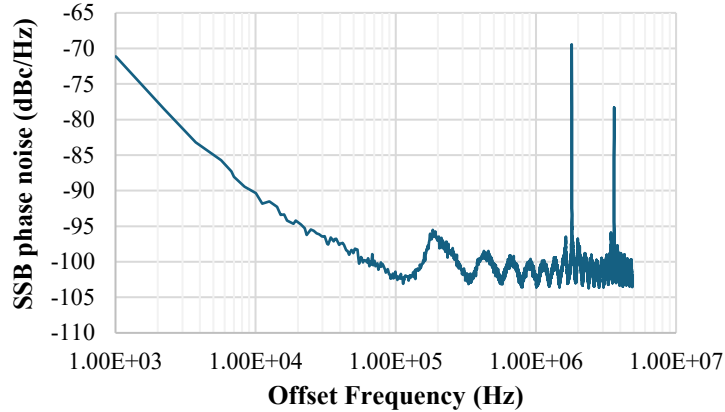

Fig. 5. SSB phase noise of the generated V-band signal.

A comparison of the parameters between the proposed OEO system and the electrical signal currently used as an intermediate signal for LO generation in Earth observation is shown in Table 2. As expected, the OEO system has much lower phase noise than the electrical signal. The noise floor measurement of OEO system was limited by the noise floor of ESA, and the actual value may in fact be much lower than -115 dBm/Hz. The frequency stability of the OEO is measured at 30 KHz for a 10-minute time interval. This measurement was taken in laboratory conditions with room temperature of 25°C. No temperature control for the separate devices was used except for the SOA which has a dedicated Peltier module.

As far as the power consumption of the OEO system is concerned, it is approximately 31 W with the vast majority of it used for the operation of the PA (around 23 W). As for the size of the OEO, it is can in principle be approximately 300 mm by 250 mm by 100 mm, provided that compact fiber spools are used [4] instead of the regular size ones and the 90-10 optical splitter for monitoring the optical signal is removed. As for the weight of the OEO, it is approximately 1975 g. The weight estimation of the OEO system does not include the weigth of the DC power sources for the LS, MZM, SOA, BPD, LNA and PA.

Table 2. Comparison of parameters between the 45.86 GHz OEO and the electrical intermediate signal currently used for LO generation in Earth observation.

| Parameter | OEO | Electrical signal |
|---|---|---|
| Output Frequency | 45.86 GHz | 45.8 GHz |
| Output power | -3.9 dBm | -5 dBm |
| Sidemode suppression ratio | 46 dBc | - |
| Phase noise | -102 dBc/Hz at 100 kHz | -75 dBc/Hz at 100 kHz |
| Noise floor | -115 dBm/Hz[1] | -159 dBm/Hz |
| Frequency stability[2] | 30 kHz | <50 kHz |

[1] The noise floor measurement was limited by the noise floor of the ESA.

[2] Frequency stability for a 10-minute time interval.

## 4. CONCLUSION

An optoelectronic oscillator (OEO) producing a signal at 45.86 GHz was demonstrated experimentally and its performance was evaluated in terms of sidemode suppression ratio, phase noise and frequency stability. The output frequency and performance of the OEO makes it a promising candidate to substitute or supplement the system currently used for LO generation in Earth observation missions of ESA. However, more research is needed on the long-term stability and operation of OEO in the Space environment to ensure that it satisfies all necessary requirements, including those related to size, weight, and power (SWaP), for use in MWS, MWI, ICI or future ESA missions.

## ACKNOWLEDGEMENTS

This work is part of the Photonic Generation of Sub-THz Signals using Optoelectronic Oscillator (Sub-THz OEO) project, which is funded by the European Space Agency (ESA) and the Republic of Cyprus under the Plan for European Cooperating States (PECS).